\documentclass[12pt]{iopart}
\expandafter\let\csname equation*\endcsname\relax
\expandafter\let\csname endequation*\endcsname\relax

\usepackage{iopams,graphicx,amsmath,amssymb,bm}
\begin{document}

\title{Absolute negative mobility in evolution}

\author{Masahiko Ueda}
\address{Department of Systems Science, Kyoto University, Kyoto, 606-8501, Japan}
\ead{ueda.masahiko.5r@kyoto-u.ac.jp}

\begin{abstract}
We investigate a population-genetic model with a temporally-fluctuating sawtooth fitness landscape.
We numerically show that a counter-intuitive behavior occurs where the rate of evolution of the system decreases as selection pressure increases from zero.
This phenomenon is understood by analogy with absolute negative mobility in particle flow.
A phenomenological explanation about the direction of evolution is also provided.
\end{abstract}

\vspace{2pc}

\maketitle

\section{Introduction}
\label{sec:intro}
Evolution of biological systems generally occurs in unsteady environments \cite{Bel2010}.
For example, climate change leads to selection of different genotypes \cite{BBOSP2014}.
Accordingly, population genetic theories in unsteady environments recently attract much attention \cite{KusLei2005,KNA2007,MusLas2008,MusLas2010,GPW2010,LeiKus2010,RivLei2011,AAG2014,PatKlu2015,CGJD2015,HLGM2016,SkaKus2016,SugKob2017,KobSug2017,MMRW2017,WanDai2019}.
Especially, in Refs. \cite{KusLei2005,RivLei2011,KobSug2017}, adaptation of population to fluctuating environments is studied in the context of information sensing.
In Refs. \cite{MusLas2010,LeiKus2010,SugKob2017}, concepts of non-equilibrium statistical mechanics are applied to population genetics.
It has been known that population genetic theory of infinite-size non-interacting populations in fluctuating environments is formally equivalent to equilibrium statistical mechanics of directed polymers \cite{KLG2006}, where the time evolution equation of population size of each genotype is mathematically equivalent to transfer matrix equation of directed polymers, and effective fitness of population corresponds to free energy of polymers.
In Refs. \cite{GPW2010,PatKlu2015,SkaKus2016,MMRW2017,WanDai2019}, transition between various types of response to environments have been reported.
In addition, there are also many studies on population genetic theory of finite-size population in specific fluctuating environments \cite{MusLas2008,OTN2011,AAG2014,CGJD2015,HLGM2016}.
In particular, it was found that temporally varying environments can speed up evolution \cite{KNA2007}.
Finiteness of population size causes genetic drift \cite{HarCla1997,SelHir2005}, and various behaviors can be observed.

One of the purpose of population genetic theory is clarifying the relation between the speed of evolution $v$ and the strength of selection pressure $s$ \cite{PSK2010}, where $v$ is defined as the rate of increase in populaton-averaged logarithmic fitness divided by $s$.
In evolution in a static environment, it is naively expected that stronger selection leads to faster evolution, that is,
\begin{eqnarray}
 \frac{dv}{ds} &>& 0 \qquad (\forall s).
 \label{eq:v-s_normal}
\end{eqnarray}
In fact, this is true for evolution driven by mutation in a smooth static fitness landscape \cite{Gil2004}.
However, a previous study revealed that this is not always the case, and stronger selection pressure can slow down evolution driven by recombination and migration even if a fitness landscape is static and smooth \cite{UTK2017}, that is,
\begin{eqnarray}
 \frac{dv}{ds} &<& 0 \qquad (\exists s).
 \label{eq:v-s_NDM}
\end{eqnarray}
Such behavior can also be observed for evolution driven by mutation in a rugged static fitness landscape.
It should be noted that this phenomenon can be understood by analogy with negative differential mobility in non-equilibrium particle flow, when we regard the speed of evolution and the strength of selection pressure as particle current and driving force, respectively.
In studies of particle flow, there is more dramatic phenomenon, called absolute negative mobility (ANM) \cite{RKVH1999,RVK1999,BPVR2000,MDW2001,CleVan2001,CleVan2002,ERH2002a,ERH2002b,CleVan2003,HMSR2004,MKTLH2007,HMSS2010,DMP2018}, which attracts much interest in the context of Brownian ratchet \cite{Rei2002}.
In ANM, direction of particle current becomes opposite to the direction of driving force even when driving force is infinitely small.
ANM has been observed in coupled stochastic equations \cite{RKVH1999,RVK1999,BPVR2000,MDW2001,CleVan2001}, non-Markovian random walk \cite{CleVan2002}, a single Brownian particle \cite{ERH2002a,ERH2002b,CleVan2003,HMSR2004}, an underdamped Brownian particle \cite{MKTLH2007}, and a Brownian elliptic disk \cite{HMSS2010}.
ANM occurs even around equilibrium states \cite{DMP2018}.
A natural question is whether there is a population genetic phenomenon corresponding to ANM, that is,
\begin{eqnarray}
 \left. \frac{dv}{ds} \right|_{s=+0} &<& 0.
\end{eqnarray}

In this study, we propose a population genetic model which exhibits a phenomenon similar to ANM.
In this model, we consider an oscillating environment where favorable genotype periodically changes, as seasonal variations in climate.
We numerically show that the rate of evolution of the system decreases as selection pressure increases from zero.

We remark that the phenomenon reported in this paper is different from other biological ratchets.
In population genetics, process called Muller's ratchet is known \cite{Mul1964,Hai1978,RWC2003,RBW2008,NehShr2012,TKK2014,OtwKru2014}.
In Muller's ratchet, deleterious mutations are accumulated in asexual populations via genetic drift, and this mechanism is considered to be related to one of the origins of sexual reproduction.
However, in Muller's ratchet, evolution in a static environment is considered, and, in particular, although the speed of evolution $v$ is negative, Eq. (\ref{eq:v-s_normal}) still holds (that is, stronger selection prevents deleterious mutations from fixing).
Therefore, Muller's ratchet is completely different from ANM-like behavior in this paper.
Another well-known ratchet in evolutionary biology is neutral evolutionary ratchet \cite{LAKDG2011}, which is one of the scenarios about why cells obtained complexity.
In this context, the word ``ratchet'' simply means unidirectional behavior, and it is not directly related to Brownian ratchet.
Furthermore, another ratchet-like behavior was recently reported \cite{DDST2015}, where evolutionary behavior of population with a rugged fitness landscape in a fluctuating environment is investigated.
Although this situation is similar to that of our study, the result is opposite:
In the setup of Ref. \cite{DDST2015}, fluctuation of environments enhances the probability that population reaches the fittest genotypes, whereas in our setup, fitness decreases in ANM-like region.
Furthermore, the main focus of our paper is $s$ dependence of $v$.
Therefore, to the best of our knowledge, the result similar to that of our paper has not been reported yet in the context of population genetics.

The paper is organized as follows.
In section \ref{sec:model}, we introduce a population genetic model which describes evolution in an oscillating environment.
In section \ref{sec:preliminaries}, we provide numerical results when an environment is static.
In section \ref{sec:results}, we provide numerical results for an oscillating environment, which is similar to ANM.
An intuitive explanation is also provided in the section.
In section \ref{sec:analysis}, we provide a phenomenological explanation why evolution to the direction decreasing fitness occurs in this model.
In section \ref{sec:discussion}, we investigate whether ANM-like behavior also occurs when the rate of evolution is defined by fitness flux \cite{MusLas2010}.
The effect of double mutation is also studied in the section.
Section \ref{sec:conclusion} is devoted to concluding remarks.

\section{Model}
\label{sec:model}
We consider a population-genetic model with $N$ asexual individuals.
Genotype of individual $j\in \left\{ 1, \cdots, N \right\}$ is described as $g_j$.
We consider the situation that genotype space is discrete and one-dimensional, that is, $g_j\in \mathbb{Z}$.
The time evolution of this system consists of two steps, that is, selection and mutation.
We first define a function $\phi$ as
\begin{eqnarray}
 \phi(g) &\equiv& \left\{
 \begin{array}{ll}
 \phi_1 \frac{g}{2} + \frac{\phi_2}{2} &\quad (g=2m) \\
 \phi_1 \frac{g-1}{2} - \frac{\phi_2}{2} &\quad (g=2m+1),
 \end{array}
 \right.
\end{eqnarray}
where $m\in \mathbb{Z}$ and $\phi_1$ and $\phi_2$ are some constants with $\phi_1>0$ and $\phi_2>0$.
We assume that an environment is fluctuating and the fitness of genotype $g$ at generation $t\in \mathbb{Z}$ is described as
\begin{eqnarray}
 W_t(g) &=& \left\{
 \begin{array}{ll}
 e^{s_0 \phi(g)} \times e^{s \phi(g)} &\quad ((t \mod \tau) < \frac{\tau}{2}) \\
 e^{-s_0 \phi(g)} \times e^{s \phi(g)} &\quad ((t \mod \tau) \geq \frac{\tau}{2}),
 \end{array}
 \right.
\end{eqnarray}
where $s_0>0$ and $\tau$ is some positive even number describing a period of environmental oscillation.
The parameter $s$ corresponds to effective selection pressure because geometric mean of fitness over one period of environmental oscillation is $e^{s \phi(g)}$.
We also call $e^{s \phi(g)}$ effective fitness.
In selection step, $N$ individuals in the next generation are independently sampled according to the probability distribution
\begin{eqnarray}
 P(j) &=& \frac{W_t(g_j)}{\sum_{k=1}^N W_t(g_k)}.
 \label{eq:WF}
\end{eqnarray}
In mutation step, genotype of each individual changes with transition probability
\begin{eqnarray}
 T\left( g|g^\prime \right) &=& \left( 1-2\mu \right) \delta_{g,g^\prime} + \mu \delta_{g,g^\prime+1} + \mu \delta_{g,g^\prime-1}.
 \label{eq:transitionprob}
\end{eqnarray}
When $s>0$, effective fitness $e^{s \phi(g)}$ is greater for larger $g$, and it is simply expected that the system evolves toward larger $g$ on average.

\begin{figure}[t]
\includegraphics[clip, width=8.0cm]{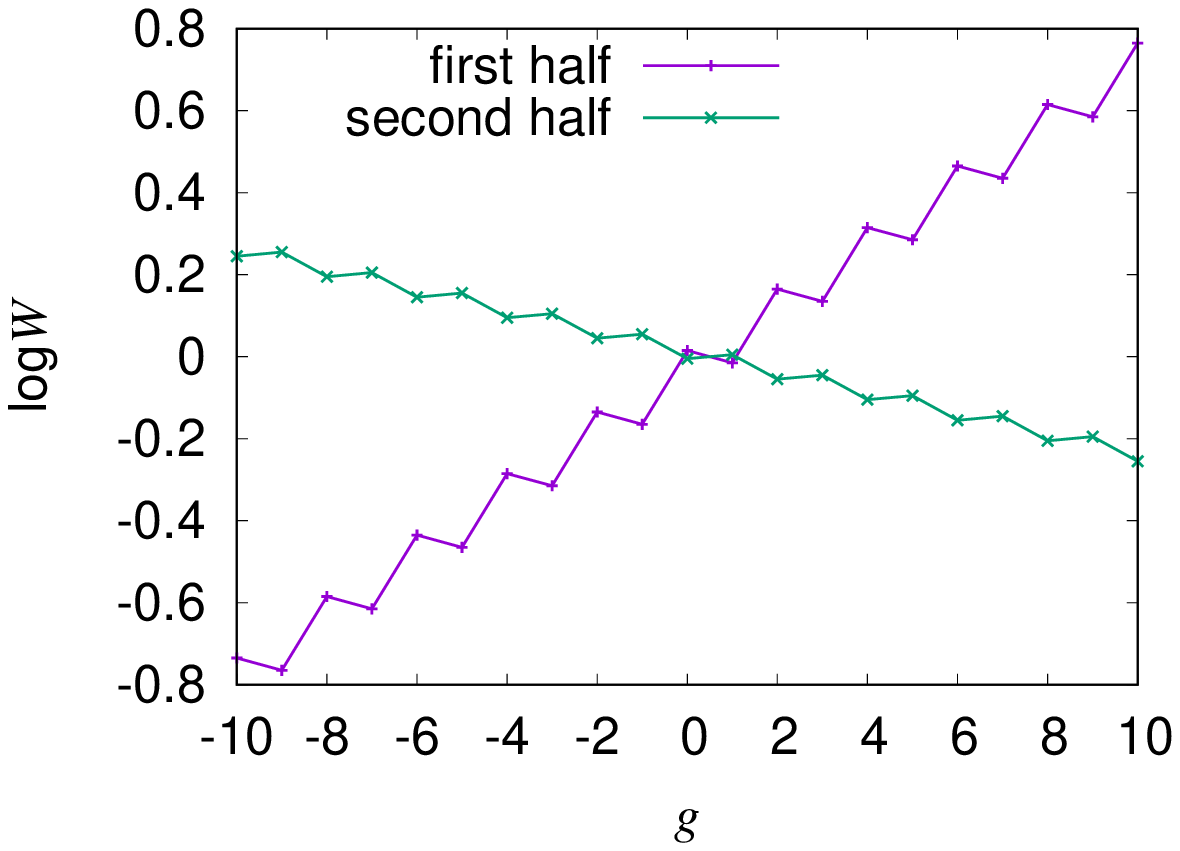}
\includegraphics[clip, width=8.0cm]{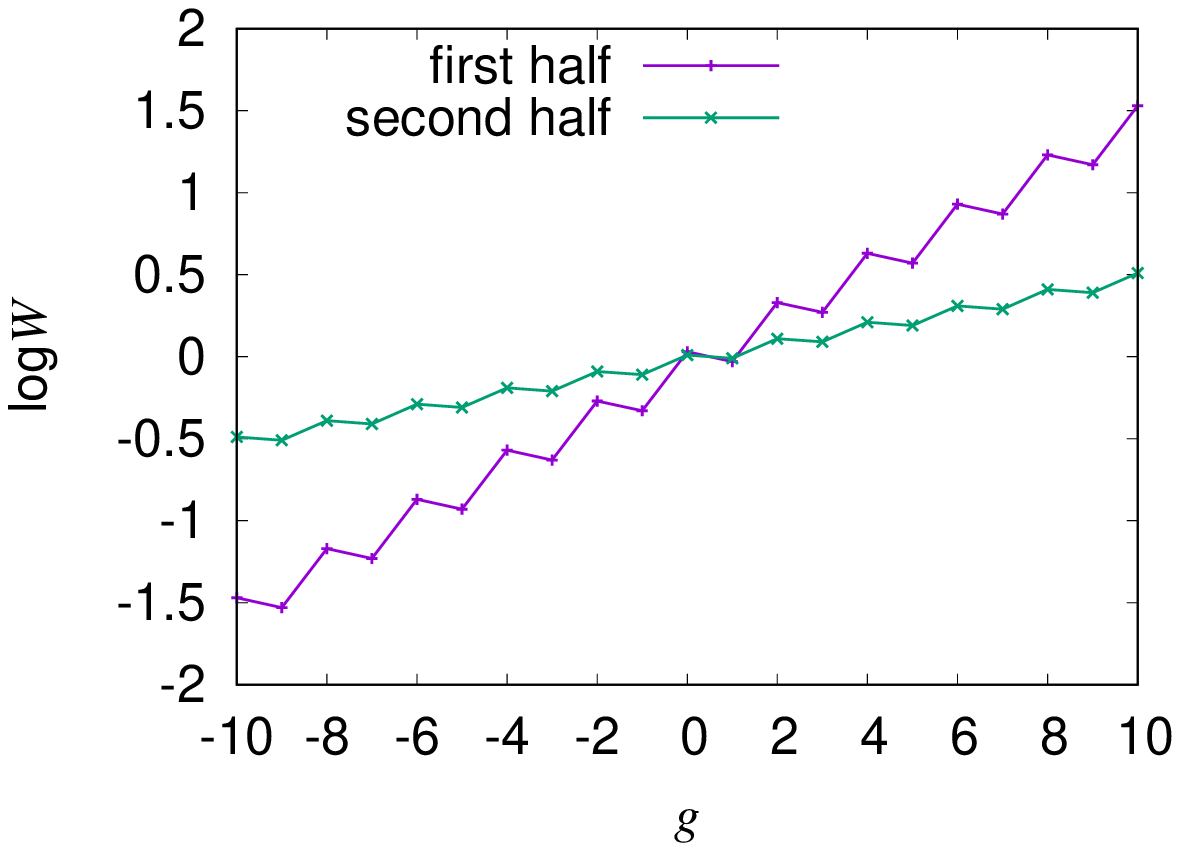}
\caption{Logarithm of fitness landscape, $\log W_t(g)$, for $(s,s_0)=(0.5,1.0)$ (left) and $(s,s_0)=(2.0,1.0)$ (right). Arithmetic mean of the two graphs in each figure is logarithm of the effective fitness, that is, $s\phi(g)$.}
\label{fig:landscape}
\end{figure}

We set parameters as $\mu=10^{-2}$, $\phi_1=0.1$, $\phi_2=0.02$ and $\tau=8$, and investigate behavior for various $s_0$ and $N$.
We display examples of fitness landscape $W_t(g)$ in Fig. \ref{fig:landscape}.
A fitness landscape is rugged for both halves of one period $\tau$.
We can see that when $s<s_0$, evolution toward smaller $g$ is possible in the second half of a period $\tau$.
In contrast, when $s>s_0$, fitness is always greater for larger $g$.
Because the latter case leads to trivial behavior, we mainly focus on the former case $s<s_0$.
Below $\left\langle \cdots \right\rangle$ describes both population average and ensemble average.
Ensemble average is calculated by using $10000$ realizations.
The initial condition is $g_j=0$ for all $j$.

\section{Preliminaries}
\label{sec:preliminaries}
Because effective fitness $e^{s \phi(g)}$ is greater for larger $g$ when we ignore bumps, the quantity
\begin{eqnarray}
 v &=& \frac{d \left\langle g \right\rangle}{dt}.
\end{eqnarray}
can be regarded as the speed of evolution, instead of $d\left\langle \phi(g) \right\rangle/dt$.

\begin{figure}[t]
\includegraphics[clip, width=8.0cm]{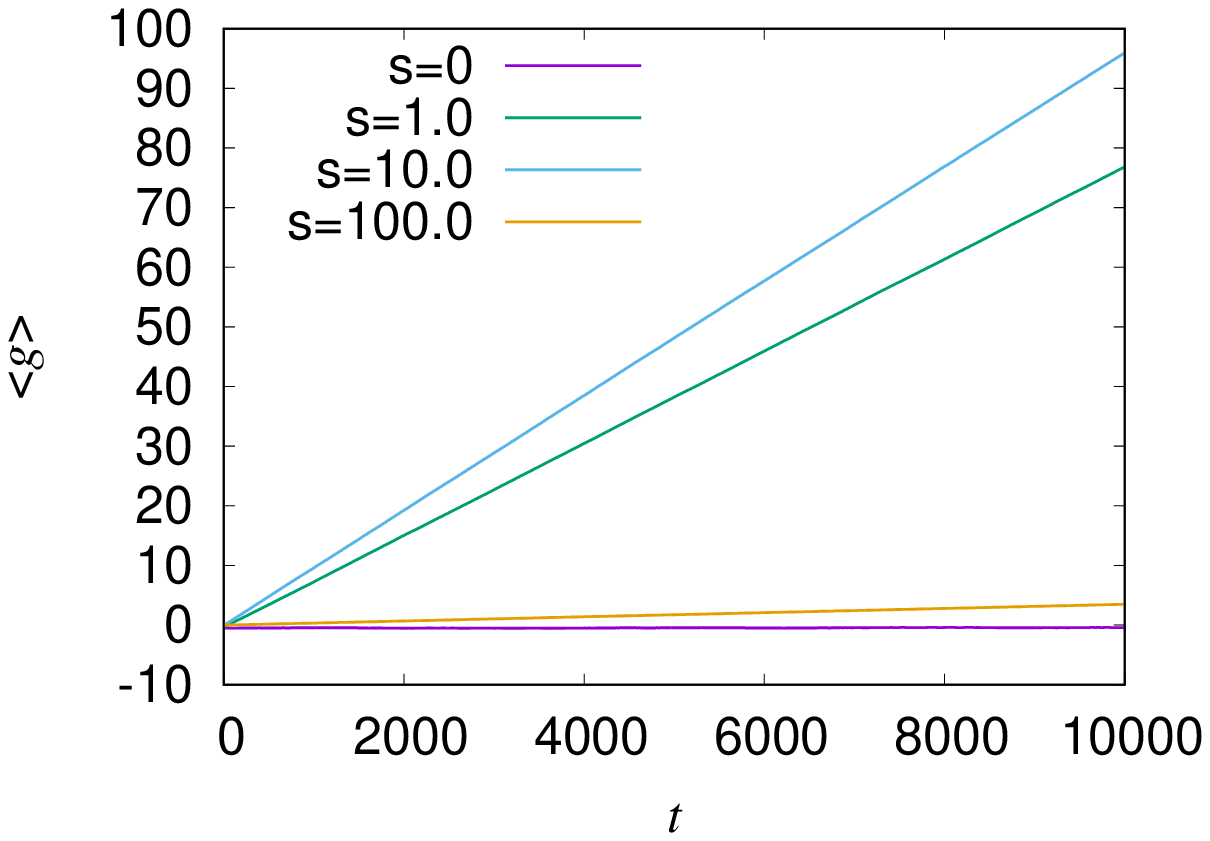}
\includegraphics[clip, width=8.0cm]{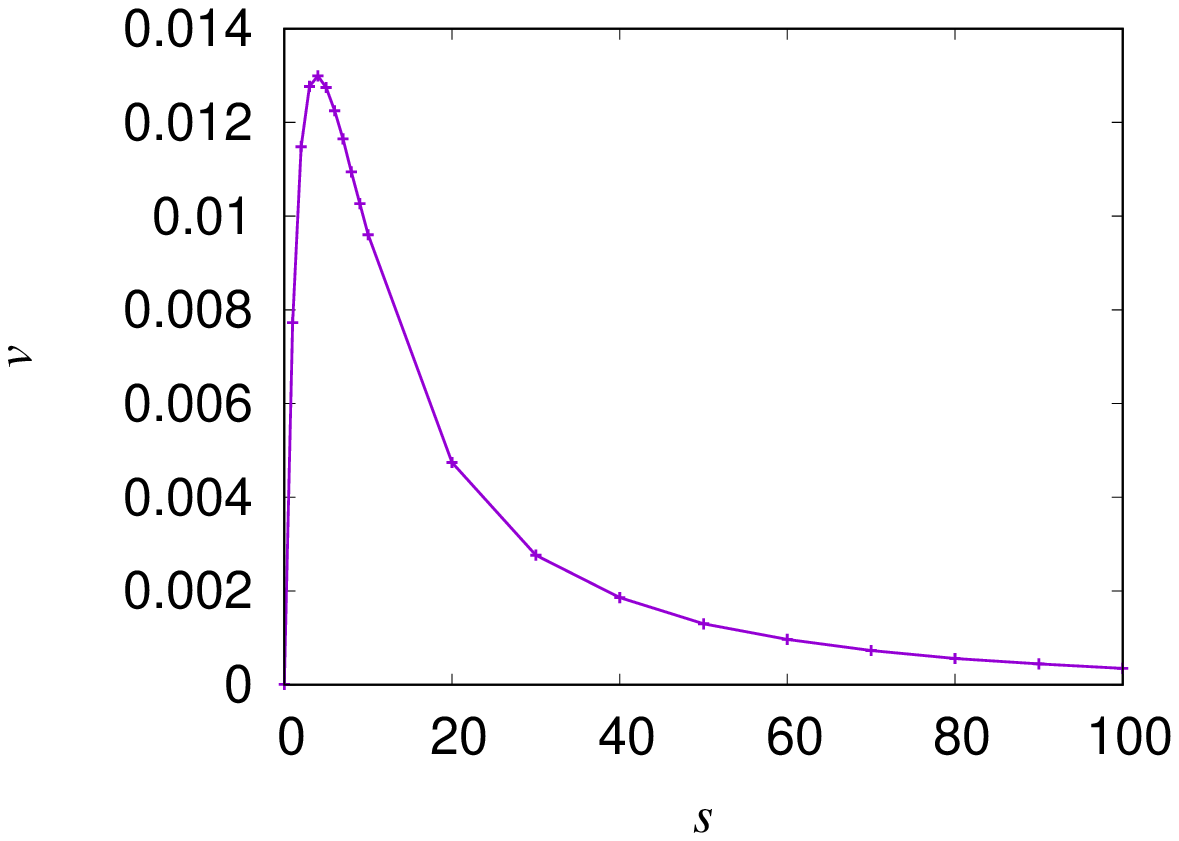}
\caption{(Left) The time evolution of $\left\langle g \right\rangle$ for $(N, s_0)=(10,0)$. (Right) $s$ dependence of $v$ for $(N, s_0)=(10,0)$.}
\label{fig:s-v_10_0}
\end{figure}

Before studying the above model, we first investigate the case $s_0=0$, that is, an environment is static.
In the left side of Fig. \ref{fig:s-v_10_0}, we have displayed the time evolution of $\left\langle g \right\rangle$ at $(N, s_0)=(10,0)$.
We can see that $\left\langle g \right\rangle$ linearly increases with $t$ for $s>0$.
We also plot the slope of $\left\langle g \right\rangle$, which is calculated by fitting a linear equation $vt+b$ to the graph by the least squares method, in the right side of Fig. \ref{fig:s-v_10_0}.
We can see that $v$ is not monotonic function of $s$.
This behavior is similar to negative differential mobility as Eq. (\ref{eq:v-s_NDM}).
The mechanism of such behavior is as follows.
When $s$ is small, selection is moderate and individuals with larger $g$ are selected.
However, when $s$ is large enough, all individuals are trapped to a local maximum, and the escape probability from the local maximum decreases with $s$.
Therefore, behavior like negative differential mobility is observed even in static landscape case.

\section{Numerical results}
\label{sec:results}
We investigate $s$ dependence of $v$ for various $(N, s_0)$.
It should be noted that $v=0$ for $s=0$ because two directions are symmetric.
We also remark that $v\rightarrow 0$ for $s \rightarrow \infty$, since all individuals are trapped to $g=0$ (in the first half of a period) or $g=-1$ (in the second half of a period) in the limit $s \rightarrow \infty$.
Therefore, negative differential mobility-like behavior is always expected to occur for large $s$ region because of the ruggedness of a fitness landscape, as in the case $s_0=0$ (section \ref{sec:preliminaries}).
In this paper, we focus on the behavior of $v$ near $s=0$.
Naively one may expect that $v$ is an increasing function of $s$ near $s=0$ because evolution effectively selects individuals with larger fitness, that is, larger $g$.
However, we see that this is not necessarily the case.

\begin{figure}[t]
\includegraphics[clip, width=8.0cm]{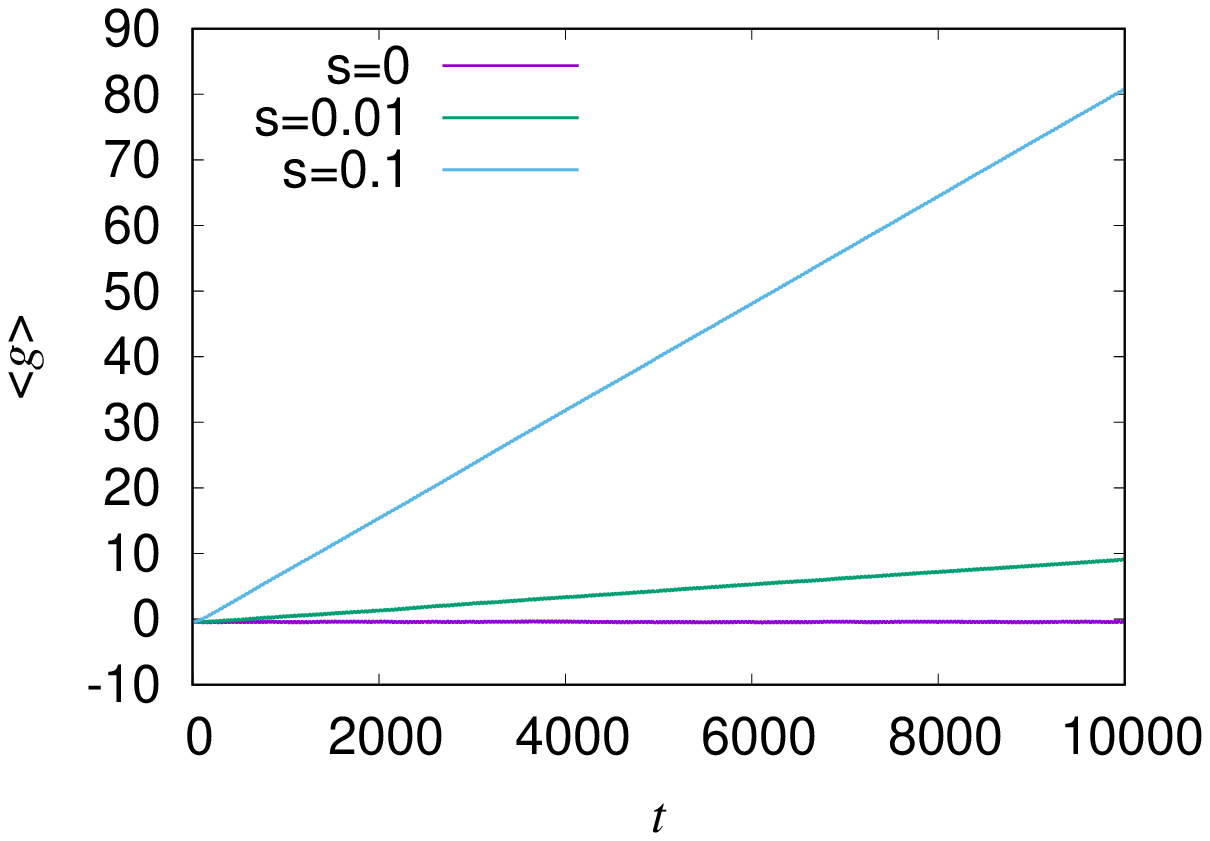}
\includegraphics[clip, width=8.0cm]{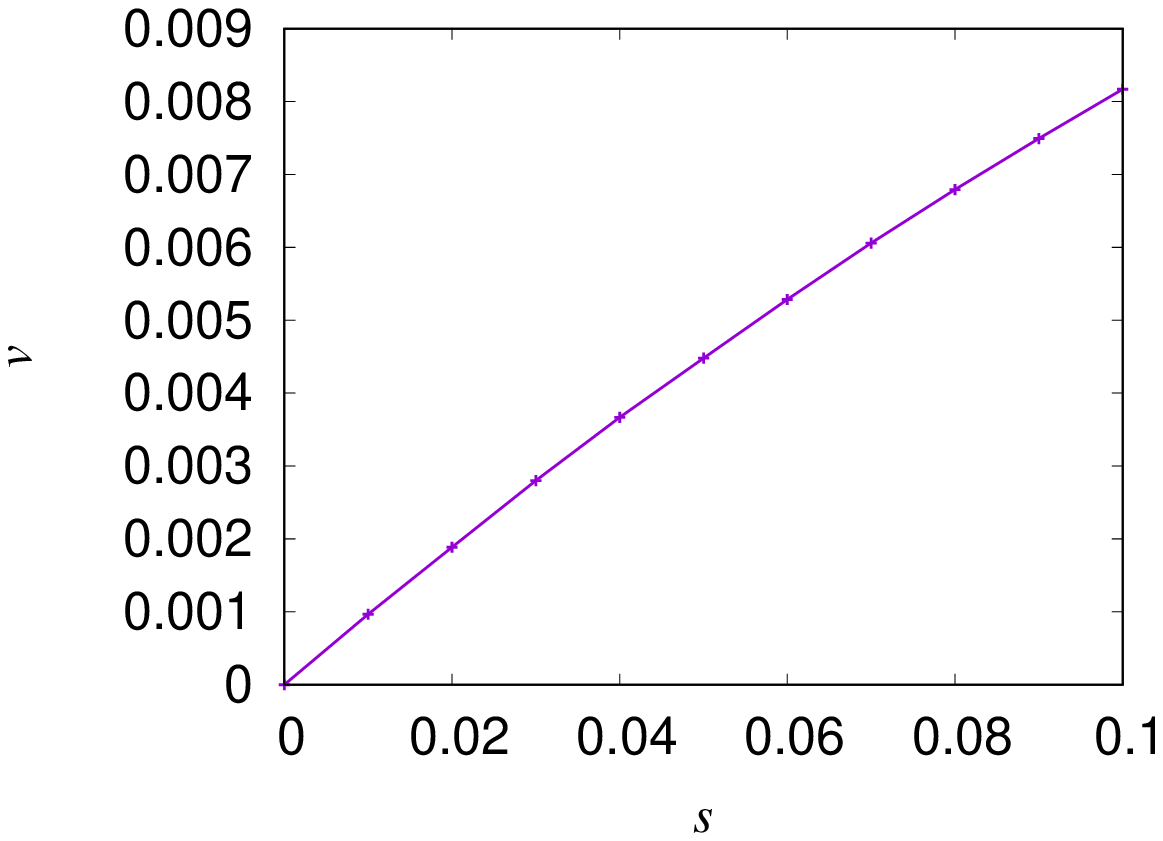}
\caption{(Left) The time evolution of $\left\langle g \right\rangle$ for $(N, s_0)=(100,1.0)$. (Right) $s$ dependence of $v$ for $(N, s_0)=(100,1.0)$.}
\label{fig:s-v_100_1}
\end{figure}

\begin{figure}[t]
\includegraphics[clip, width=8.0cm]{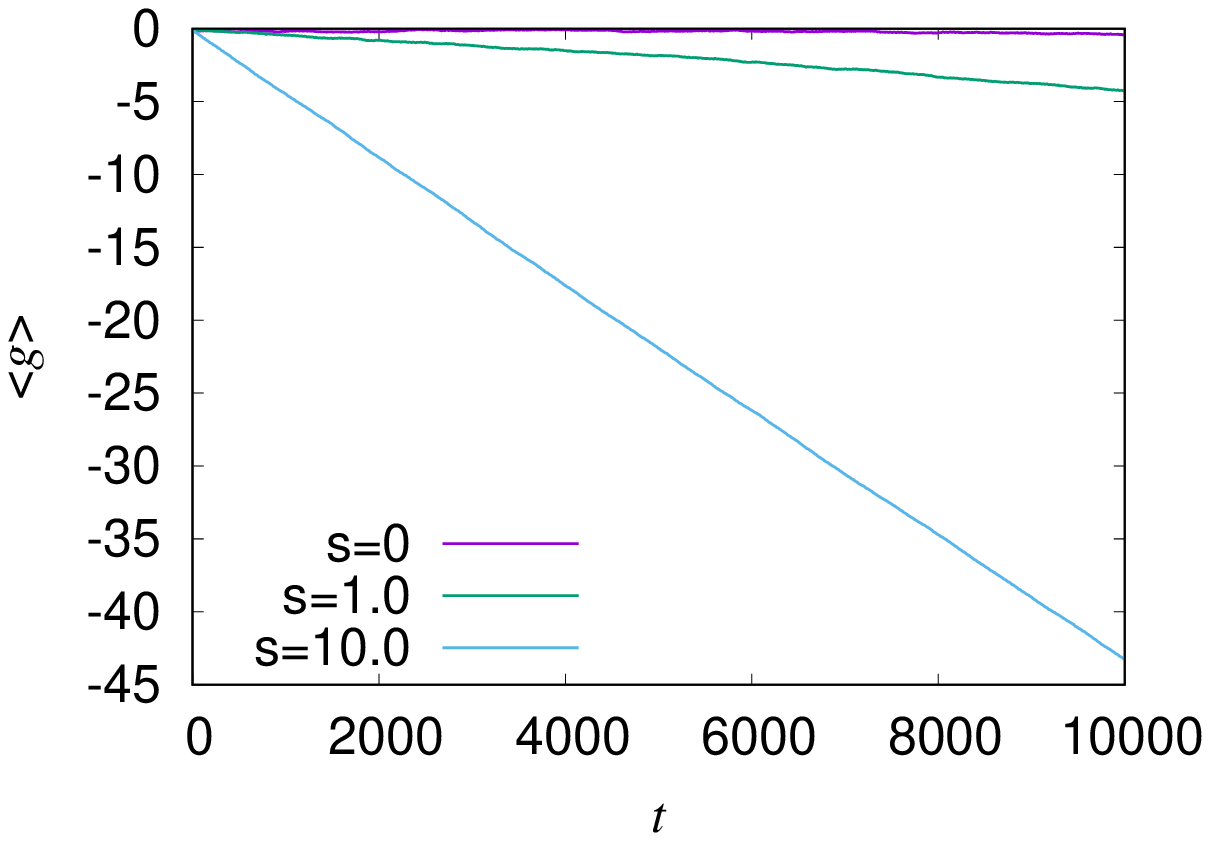}
\includegraphics[clip, width=8.0cm]{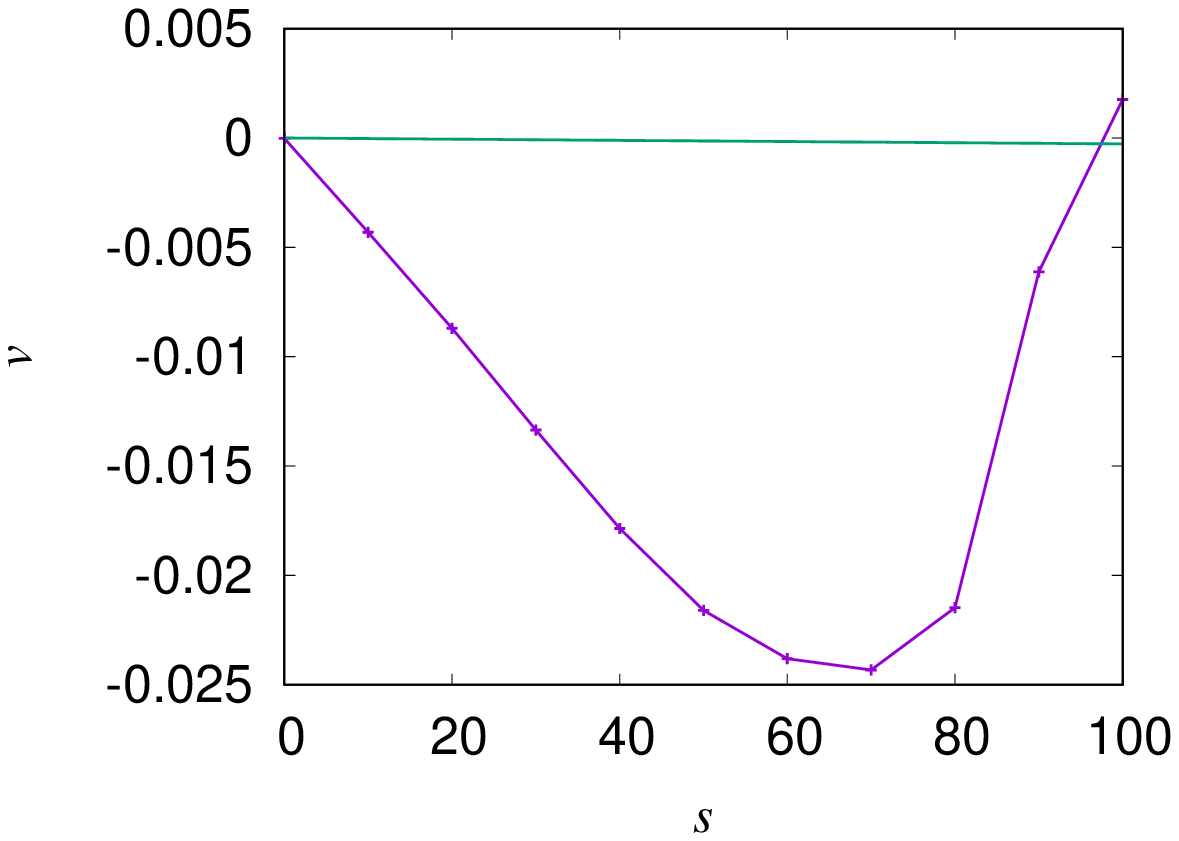}
\caption{(Left) The time evolution of $\left\langle g \right\rangle$ for $(N, s_0)=(10,100.0)$. (Right) $s$ dependence of $v$ for $(N, s_0)=(10,100.0)$. The straight line is the theoretical prediction (\ref{eq:v_theory}) for small $s$.}
\label{fig:s-v_10_100}
\end{figure}

In the left side of Fig. \ref{fig:s-v_100_1}, we have displayed the time evolution of $\left\langle g \right\rangle$ for $(N, s_0)=(100,1.0)$.
We can see that $\left\langle g \right\rangle$ linearly increases with $t$ for $s>0$.
We also plot the slope of $\left\langle g \right\rangle$, which is calculated by fitting a linear equation $vt+b$ to the graph by the least squares method, in the right side of Fig. \ref{fig:s-v_100_1}.
We can see that $v$ is increasing function of $s$ near $s=0$.
In contrast, in the left side of Fig. \ref{fig:s-v_10_100}, we have displayed the time evolution of $\left\langle g \right\rangle$ for $(N, s_0)=(10,100.0)$.
We can see that $\left\langle g \right\rangle$ linearly decreases with $t$ for $s>0$.
We also plot the slope of $\left\langle g \right\rangle$ in the right side of Fig. \ref{fig:s-v_10_100}.
We can see that $v$ is negative for small $s$, which is ANM-like behavior.

\begin{figure}[t]
\includegraphics[clip, width=8.0cm]{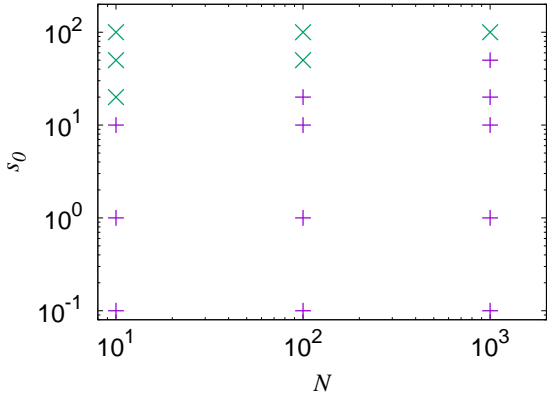}
\caption{The $N$-$s_0$ phase diagram. The purple symbol $+$ represents normal phase and the green symbol $\times$ represents ANM phase.}
\label{fig:phase}
\end{figure}

We provide the $N$-$s_0$ phase diagram in Fig. \ref{fig:phase}.
In this figure, ``normal phase'' describes the parameter region where no ANM-like behavior is observed (as in the right side of Fig. \ref{fig:s-v_100_1}).
In contrast, ``ANM phase'' describes the parameter region where ANM-like behavior is observed (as in the right side of Fig. \ref{fig:s-v_10_100}). 
We find that large enough $s_0$ is necessary for ANM-like behavior.
In addition, population size $N$ should be small enough for ANM-like behavior to occur.

An intuitive picture of this phenomenon is as follows.
As we can see in the left side of Fig. \ref{fig:landscape}, the absolute value of average slope of a fitness landscape for the first half of a period is greater than that for the second half of a period.
However, the ruggedness of the former is also greater than that of the latter.
When selection pressure is large enough, trapping in local maxima occurs.
Individuals can escape trapping by genetic drift, which occurs for small population size, and this effect is stronger for lower ruggedness case, that is, the second half of a period.
Therefore, evolution to smaller $g$ occurs for small $N$ and large $s_0$.
It should be noted that this mechanism is similar to ANM in particle flow in an oscillating external field \cite{ERH2002a}.

We can also interpret this phenomenon in the context of the results for static landscape case in section \ref{sec:preliminaries}.
When $\tau$ is large enough, the speed of evolution $v$ is roughly estimated as
\begin{eqnarray}
 v &\simeq& \frac{1}{2} \left[ v_\mathrm{static}\left( s_0+s \right) - v_\mathrm{static}\left( s_0-s \right) \right],
 \label{eq:v_diff}
\end{eqnarray}
where $v_\mathrm{static}\left( s \right)$ is the speed of evolution in a static environment when selection pressure is $s$.
As we saw in section \ref{sec:preliminaries}, $v_\mathrm{static}\left( s \right)$ exhibits negative differential mobility-like behavior.
Therefore, for large enough $s_0$ and small $s$, $v_\mathrm{static}\left( s_0-s \right)>v_\mathrm{static}\left( s_0+s \right)$ holds, and $v$ becomes negative.

\section{Analysis}
\label{sec:analysis}
We consider weak-mutation strong-selection region, where $s_0\phi_1 \gg 1$, $s_0\phi_2 \gg 1$ and $\mu N \ll 1$.
In this region, population is localized to one local maximum in each generation, and the probability that evolution occurs to each direction is dominated by the tunneling probability through a local minimum by mutation.
Because we are interested in the behavior of $v$ around $s\simeq 0$, we assume that $s\ll s_0$.

Since a rough approximation for $v$ is obtained by Eq. (\ref{eq:v_diff}), we first estimate $v_\mathrm{static}(s_0)$.
In Ref. \cite{WeiCha2005}, the rate of evolution in a static rugged fitness landscape with one valley genotype and one escape genotype in weak-mutation strong-selection region was estimated.
For our case, the result is
\begin{eqnarray}
 v_\mathrm{static}(s_0) &\simeq& \mu \frac{N\mu}{1-e^{-s_0\phi_2}} \frac{1-e^{-2\frac{1}{2}s_0\phi_1}}{1-e^{-2N\frac{1}{2}s_0\phi_1}},
 \label{eq:v_static_theory}
\end{eqnarray}
where we have used the fact that our model is haploid.
The second factor comes from mutation-selection balance between an original genotype and a valley genotype \cite{CroKim1970}, and the third factor is the fixation probability of an escape genotype \cite{Kim1962}.
This expression means that, because of mutation-selection balance, $N\mu/(1-e^{-s_0\phi_2})$ individuals are in a valley on average, and when mutation occurs to these individuals, escape from the valley occurs.
Since we consider the situation $\phi_2<\phi_1<N\phi_1$, Eq. (\ref{eq:v_static_theory}) is approximated as
\begin{eqnarray}
 v_\mathrm{static}(s_0) &\simeq& N\mu^2 (1+e^{-s_0\phi_2}).
\end{eqnarray}
Therefore, from Eq. (\ref{eq:v_diff}) we finally obtain for small $s$
\begin{eqnarray}
 v &\simeq& - N \mu^2 e^{-s_0\phi_2} s \phi_2,
 \label{eq:v_theory}
\end{eqnarray}
which is negative.

\begin{figure}[t]
\includegraphics[clip, width=8.0cm]{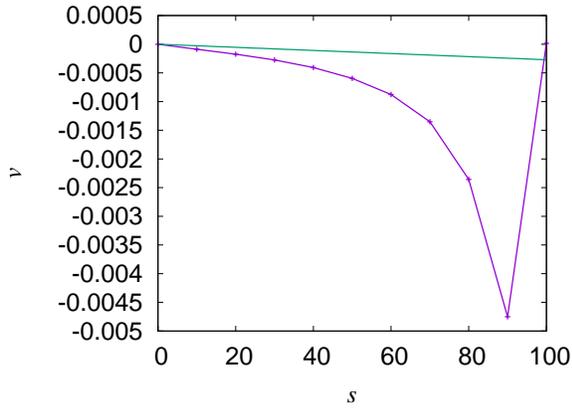}
\caption{The right-hand side of (\ref{eq:v_diff}) for $(N, s_0)=(10,100.0)$. The straight line is the theoretical prediction (\ref{eq:v_theory}) for small $s$.}
\label{fig:s-v_10_100_largeperiod}
\end{figure}

In the right side of Fig. \ref{fig:s-v_10_100}, we display the theoretical prediction (\ref{eq:v_theory}).
We find that Eq. (\ref{eq:v_theory}) does not quantitatively estimate our numerical results.
One possible cause of this discrepancy is that the period of oscillation $\tau$ is small in our numerical simulation and the oscillation increases the escape probability.
In order to check this possibility, we numerically calculate the right-hand side of (\ref{eq:v_diff}) directly, which corresponds to the case $\tau=10000$ and the phase of an environment is given at random.
We display the numerical results in Fig. \ref{fig:s-v_10_100_largeperiod}.
We can find that $|v|$ is smaller compared with that in Fig. \ref{fig:s-v_10_100} and the numerical results are more consistent with the theoretical prediction (\ref{eq:v_theory}).
We emphasize that ANM-like behavior still occurs for this case.
Therefore, we conclude that the smallness of $\tau$ enhances escape from local maxima.

Although we here consider only weak-mutation strong-selection region, our numerical results (Fig. \ref{fig:phase}) suggest that ANM-like behavior is observed in broader parameter region.

\section{Discussion}
\label{sec:discussion}
\subsection{Fitness flux}
In Ref. \cite{MusLas2010}, it was pointed out that, in fluctuating environments, the rate of evolution is not simply the increase of fitness, but fitness flux has to be considered, which satisfies integral fluctuation theorem \cite{Sei2012}.
In our discrete-time setup, the fitness flux is defined as
\begin{eqnarray}
 \Phi(t) &\equiv& \sum_{t^\prime=0}^t \sum_g \Delta x_{t^\prime}(g) \log W_{t^\prime}(g),
\end{eqnarray}
where $x_t(g)$ is the frequency of genotype $g$ at generation $t$, and $\Delta x_t(g)\equiv x_{t+1}(g) - x_t(g)$.
We would like to investigate whether the rate of increase of the fitness flux
\begin{eqnarray}
 V &\equiv& \frac{d \left\langle \Phi \right\rangle}{d t}
\end{eqnarray}
is an increasing function of the selection pressure $s$ or not.

\begin{figure}[t]
\includegraphics[clip, width=8.0cm]{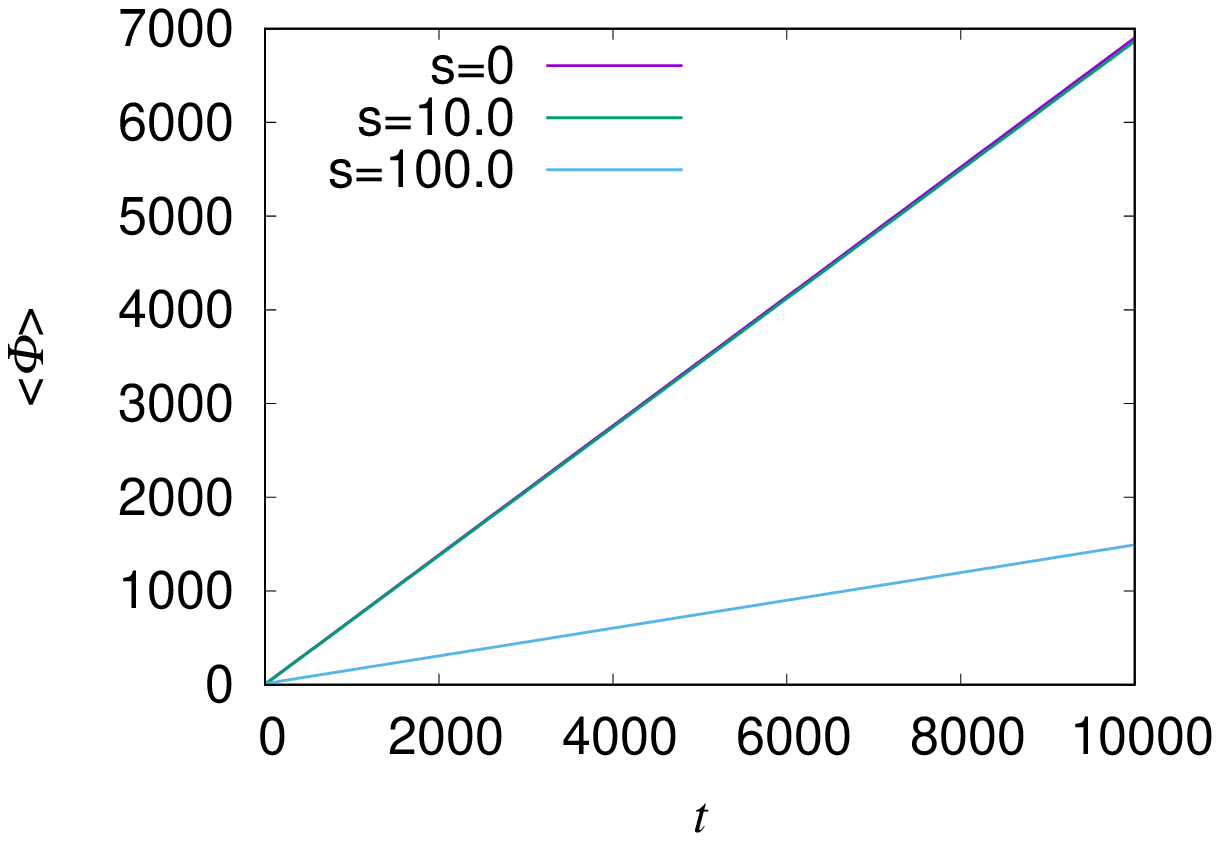}
\includegraphics[clip, width=8.0cm]{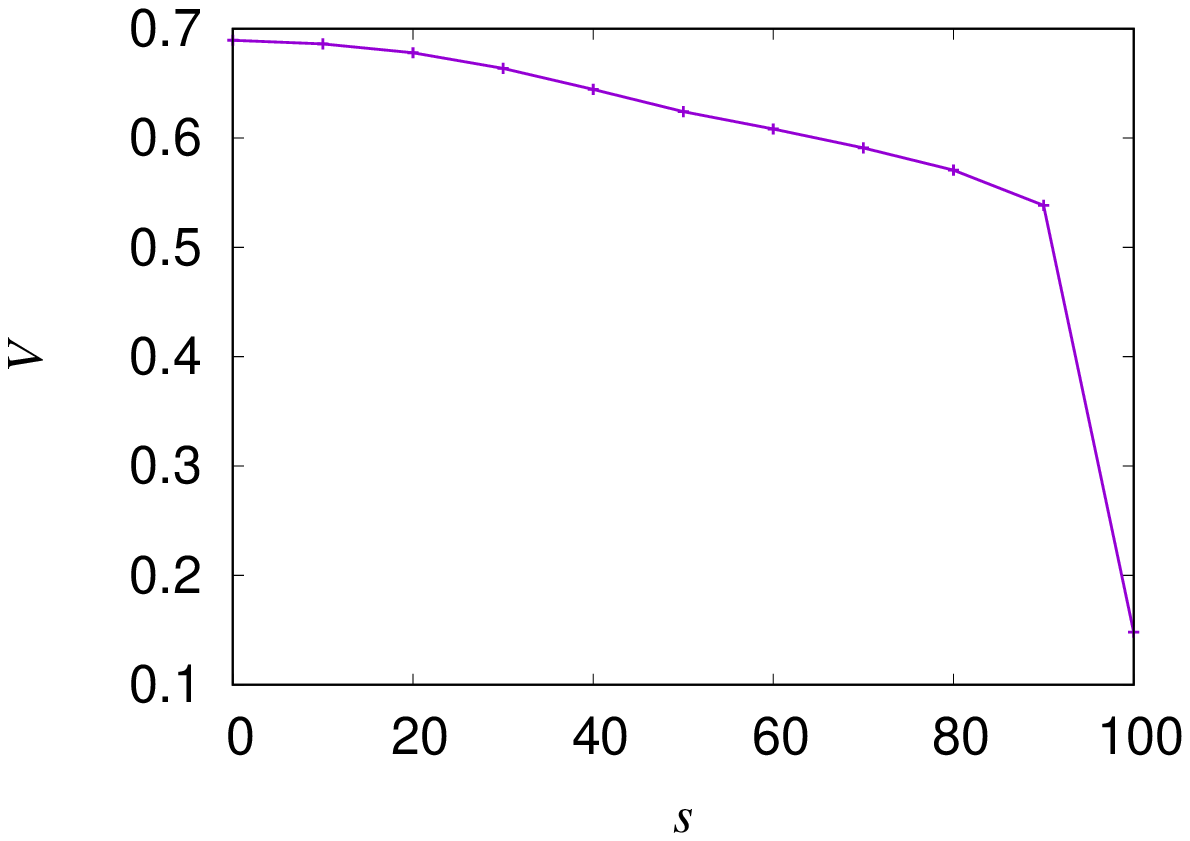}
\caption{(Left) The time evolution of $\left\langle \Phi \right\rangle$ for $(N, s_0)=(10,100.0)$. (Right) $s$ dependence of $V$ for $(N, s_0)=(10,100.0)$.}
\label{fig:s-V_10_100}
\end{figure}

In the left side of Fig. \ref{fig:s-V_10_100}, we display the time evolution of $\left\langle \Phi \right\rangle$ at $(N, s_0)=(10,100.0)$ for various $s$.
We can find that $\left\langle \Phi \right\rangle$ is an increasing function of $t$, and in particular, linear in $t$.
It should be noted that $\left\langle \Phi \right\rangle$ has positive slope even when $s=0$; this is because adaptation to an environment occurs for each half of a period.
In the right side of Fig. \ref{fig:s-V_10_100}, we display $s$ dependence of $V$, which is calculated by fitting a linear equation $Vt+b$ to the graph of $\left\langle \Phi \right\rangle$ by the least squares method.
We observe that $V$ is a monotonically decreasing function of $s$.
Therefore, we conclude that absolute negative mobility occurs even in the rate of increase of fitness flux $V$.

\subsection{Effect of double mutation}
\begin{figure}[t]
\includegraphics[clip, width=8.0cm]{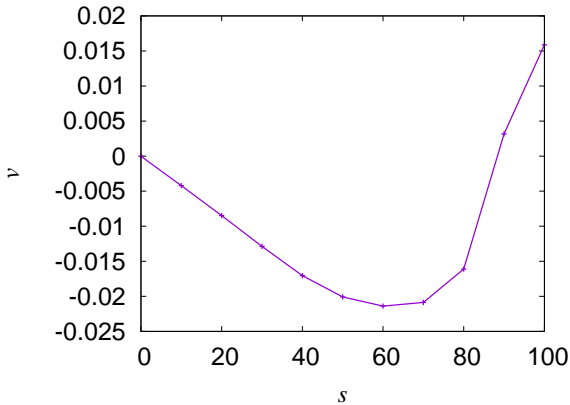}
\caption{$s$ dependence of $v$ for $(N, s_0)=(10,100.0)$ and $\mu_2=10^{-3}$.}
\label{fig:s-v_10_100_double}
\end{figure}

The ANM-like behavior reported in this paper results from trapping in local maxima.
A natural question is whether this behavior is destroyed by small probability of a double mutation, which enables individuals to escape directly from local maxima.
In order to investigate this effect, we study the same model with transition probability
\begin{eqnarray}
\fl T\left( g|g^\prime \right) &=& \left( 1-2\mu-2\mu_2 \right) \delta_{g,g^\prime} + \mu \left( \delta_{g,g^\prime+1} + \delta_{g,g^\prime-1} \right) + \mu_2 \left( \delta_{g,g^\prime+2} + \delta_{g,g^\prime-2} \right)
\end{eqnarray}
instead of Eq. (\ref{eq:transitionprob}).
We set double-mutation probability $\mu_2$ as $\mu_2=10^{-3}$.
In Fig. \ref{fig:s-v_10_100_double}, we display $s$ dependence of the speed of evolution $v$ for $(N, s_0)=(10,100.0)$.
We can see that ANM-like behavior is still observed as in Fig. \ref{fig:s-v_10_100} even if double mutation occurs with small probability.
We remark that, as $\mu_2$ further increases, ANM-like behavior will be destroyed because individuals can easily escape from local maxima.
Theoretical estimation of the effect of $\mu_2$ on $v$ is a future problem.

\section{Concluding remarks}
\label{sec:conclusion}
In this paper, we proposed a population-genetic model where the speed of evolution can decrease as selection pressure increases from zero, which has never been reported elsewhere.
This model describes evolution in a temporally-oscillating sawtooth fitness landscape.
We numerically showed that this behavior occurs when selection is strong enough and population size is small enough.
The mechanism of this phenomenon is similar to ANM in particle flow.
Although complete theoretical analysis has not been provided yet, we provide intuitive and phenomenological explanation about the direction of evolution in weak-mutation strong-selection region.
We also numerically showed that ANM-like behavior is observed even when the rate of evolution is defined through the fitness flux.

Although we considered a specific oscillating landscape where the direction of evolution in the first half of a period and that in the second half are opposite to each other, the biological meaning of this model is not clear.
Seasonal variations in climate may be similar to the situation considered in this paper, because favorable genotypes in two seasons are opposite to each other.
In such situation, $s\phi(g)$ can be interpreted to come from some traits which do not depend on the seasons.
Specifying a realistic situation described by this model is an important future problem.

Another future problem is whether there is a useful definition of efficiency of evolution.
Recently, in non-equilibrium statistical mechanics, a useful definition of transport efficiency was proposed \cite{DecSas2018}, where entropy production plays a significant role.
Whether similar concept of efficiency can be defined in the speed of evolution should be studied in future.

\ack
This study was supported by JSPS KAKENHI Grant Numbers JP16J00178 and JP19K21542.

\section*{References}
\bibliographystyle{iopart-num}
\bibliography{anm_bib_191018}

\providecommand{\newblock}{}
\begin{thebibliography}{10}
\expandafter\ifx\csname url\endcsname\relax
  \def\url#1{{\tt #1}}\fi
\expandafter\ifx\csname urlprefix\endcsname\relax\def\urlprefix{URL }\fi
\providecommand{\eprint}[2][]{\url{#2}}

\bibitem{Bel2010}
Bell G.
\newblock Fluctuating selection: the perpetual renewal of adaptation in
  variable environments 2010 {\em Philosophical Transactions of the Royal
  Society B: Biological Sciences\/} {\bf 365} 87--97

\bibitem{BBOSP2014}
Bergland A~O, Behrman E~L, O'Brien K~R, Schmidt P~S and Petrov D~A.
\newblock Genomic evidence of rapid and stable adaptive oscillations over
  seasonal time scales in drosophila 2014 {\em PLoS Genetics\/} {\bf 10}
  e1004775

\bibitem{KusLei2005}
Kussell E and Leibler S.
\newblock Phenotypic diversity, population growth, and information in
  fluctuating environments 2005 {\em Science\/} {\bf 309} 2075--2078

\bibitem{KNA2007}
Kashtan N, Noor E and Alon U.
\newblock Varying environments can speed up evolution 2007 {\em Proceedings of
  the National Academy of Sciences\/} {\bf 104} 13711--13716

\bibitem{MusLas2008}
Mustonen V and L{\"a}ssig M.
\newblock Molecular evolution under fitness fluctuations 2008 {\em Physical
  review letters\/} {\bf 100} 108101

\bibitem{MusLas2010}
Mustonen V and L{\"a}ssig M.
\newblock Fitness flux and ubiquity of adaptive evolution 2010 {\em Proceedings
  of the National Academy of Sciences\/} {\bf 107} 4248--4253

\bibitem{GPW2010}
Ga{\'a}l B, Pitchford J~W and Wood A~J.
\newblock Exact results for the evolution of stochastic switching in variable
  asymmetric environments 2010 {\em Genetics\/} {\bf 184} 1113--1119

\bibitem{LeiKus2010}
Leibler S and Kussell E.
\newblock Individual histories and selection in heterogeneous populations 2010
  {\em Proceedings of the National Academy of Sciences\/} {\bf 107}
  13183--13188

\bibitem{RivLei2011}
Rivoire O and Leibler S.
\newblock The value of information for populations in varying environments 2011
  {\em Journal of Statistical Physics\/} {\bf 142} 1124--1166

\bibitem{AAG2014}
Ashcroft P, Altrock P~M and Galla T.
\newblock Fixation in finite populations evolving in fluctuating environments
  2014 {\em Journal of The Royal Society Interface\/} {\bf 11} 20140663

\bibitem{PatKlu2015}
Patra P and Klumpp S.
\newblock Emergence of phenotype switching through continuous and discontinuous
  evolutionary transitions 2015 {\em Physical biology\/} {\bf 12} 046004

\bibitem{CGJD2015}
Cvijovi{\'c} I, Good B~H, Jerison E~R and Desai M~M.
\newblock Fate of a mutation in a fluctuating environment 2015 {\em Proceedings
  of the National Academy of Sciences\/} {\bf 112} E5021--E5028

\bibitem{HLGM2016}
Hufton P~G, Lin Y~T, Galla T and McKane A~J.
\newblock Intrinsic noise in systems with switching environments 2016 {\em
  Physical Review E\/} {\bf 93} 052119

\bibitem{SkaKus2016}
Skanata A and Kussell E.
\newblock Evolutionary phase transitions in random environments 2016 {\em
  Physical review letters\/} {\bf 117} 038104

\bibitem{SugKob2017}
Sughiyama Y and Kobayashi T~J.
\newblock Steady-state thermodynamics for population growth in fluctuating
  environments 2017 {\em Physical Review E\/} {\bf 95} 012131

\bibitem{KobSug2017}
Kobayashi T~J and Sughiyama Y.
\newblock Stochastic and information-thermodynamic structures of population
  dynamics in a fluctuating environment 2017 {\em Physical Review E\/} {\bf 96}
  012402

\bibitem{MMRW2017}
Mayer A, Mora T, Rivoire O and Walczak A~M.
\newblock Transitions in optimal adaptive strategies for populations in
  fluctuating environments 2017 {\em Physical Review E\/} {\bf 96} 032412

\bibitem{WanDai2019}
Wang S and Dai L.
\newblock Evolving generalists in switching rugged landscapes 2019 {\em PLoS
  computational biology\/} {\bf 15} e1007320

\bibitem{KLG2006}
Kussell E, Leibler S and Grosberg A.
\newblock Polymer-population mapping and localization in the space of
  phenotypes 2006 {\em Physical review letters\/} {\bf 97} 068101

\bibitem{OTN2011}
Otwinowski J, Tanase-Nicola S and Nemenman I.
\newblock Speeding up evolutionary search by small fitness fluctuations 2011
  {\em Journal of statistical physics\/} {\bf 144} 367

\bibitem{HarCla1997}
Hartl D~L and Clark A~G 1997 {\em Principles of population genetics\/} vol 116
  (Sinauer associates Sunderland, MA)

\bibitem{SelHir2005}
Sella G and Hirsh A~E.
\newblock The application of statistical physics to evolutionary biology 2005
  {\em Proceedings of the National Academy of Sciences\/} {\bf 102} 9541--9546

\bibitem{PSK2010}
Park S~C, Simon D and Krug J.
\newblock The speed of evolution in large asexual populations 2010 {\em Journal
  of Statistical Physics\/} {\bf 138} 381--410

\bibitem{Gil2004}
Gillespie J~H 2004 {\em Population genetics: a concise guide\/} (JHU Press)

\bibitem{UTK2017}
Ueda M, Takeuchi N and Kaneko K.
\newblock Stronger selection can slow down evolution driven by recombination on
  a smooth fitness landscape 2017 {\em PloS one\/} {\bf 12} e0183120

\bibitem{RKVH1999}
Reimann P, Kawai R, Van~den Broeck C and H{\"a}nggi P.
\newblock Coupled brownian motors: Anomalous hysteresis and zero-bias negative
  conductance 1999 {\em EPL (Europhysics Letters)\/} {\bf 45} 545

\bibitem{RVK1999}
Reimann P, Van~den Broeck C and Kawai R.
\newblock Nonequilibrium noise in coupled phase oscillators 1999 {\em Physical
  Review E\/} {\bf 60} 6402

\bibitem{BPVR2000}
Buceta J, Parrondo J, Van~den Broeck C and de~La~Rubia F.
\newblock Negative resistance and anomalous hysteresis in a collective
  molecular motor 2000 {\em Physical Review E\/} {\bf 61} 6287

\bibitem{MDW2001}
Mangioni S, Deza R and Wio H.
\newblock Transition from anomalous to normal hysteresis in a system of coupled
  brownian motors: A mean-field approach 2001 {\em Physical Review E\/} {\bf
  63} 041115

\bibitem{CleVan2001}
Cleuren B and Van~den Broeck C.
\newblock Ising model for a brownian donkey 2001 {\em EPL (Europhysics
  Letters)\/} {\bf 54} 1

\bibitem{CleVan2002}
Cleuren B and Van~den Broeck C.
\newblock Random walks with absolute negative mobility 2002 {\em Physical
  Review E\/} {\bf 65} 030101

\bibitem{ERH2002a}
Eichhorn R, Reimann P and H{\"a}nggi P.
\newblock Brownian motion exhibiting absolute negative mobility 2002 {\em
  Physical review letters\/} {\bf 88} 190601

\bibitem{ERH2002b}
Eichhorn R, Reimann P and H{\"a}nggi P.
\newblock Paradoxical motion of a single brownian particle: Absolute negative
  mobility 2002 {\em Physical Review E\/} {\bf 66} 066132

\bibitem{CleVan2003}
Cleuren B and Van~den Broeck C.
\newblock Brownian motion with absolute negative mobility 2003 {\em Physical
  Review E\/} {\bf 67} 055101

\bibitem{HMSR2004}
Haljas A, Mankin R, Sauga A and Reiter E.
\newblock Anomalous mobility of brownian particles in a tilted symmetric
  sawtooth potential 2004 {\em Physical Review E\/} {\bf 70} 041107

\bibitem{MKTLH2007}
Machura L, Kostur M, Talkner P, {\L}uczka J and H{\"a}nggi P.
\newblock Absolute negative mobility induced by thermal equilibrium
  fluctuations 2007 {\em Physical review letters\/} {\bf 98} 040601

\bibitem{HMSS2010}
H{\"a}nggi P, Marchesoni F, Savel'ev S and Schmid G.
\newblock Asymmetry in shape causing absolute negative mobility 2010 {\em
  Physical Review E\/} {\bf 82} 041121

\bibitem{DMP2018}
Cividini J, Mukamel D and Posch H.
\newblock Driven tracer with absolute negative mobility 2018 {\em Journal of
  Physics A: Mathematical and Theoretical\/} {\bf 51} 085001

\bibitem{Rei2002}
Reimann P.
\newblock Brownian motors: noisy transport far from equilibrium 2002 {\em
  Physics reports\/} {\bf 361} 57--265

\bibitem{Mul1964}
Muller H~J.
\newblock The relation of recombination to mutational advance 1964 {\em
  Mutation Research/Fundamental and Molecular Mechanisms of Mutagenesis\/} {\bf
  1} 2--9

\bibitem{Hai1978}
Haigh J.
\newblock The accumulation of deleterious genes in a population―muller's
  ratchet 1978 {\em Theoretical population biology\/} {\bf 14} 251--267

\bibitem{RWC2003}
Rouzine I~M, Wakeley J and Coffin J~M.
\newblock The solitary wave of asexual evolution 2003 {\em Proceedings of the
  National Academy of Sciences\/} {\bf 100} 587--592

\bibitem{RBW2008}
Rouzine I~M, Brunet {\'E} and Wilke C~O.
\newblock The traveling-wave approach to asexual evolution: Muller's ratchet
  and speed of adaptation 2008 {\em Theoretical population biology\/} {\bf 73}
  24--46

\bibitem{NehShr2012}
Neher R~A and Shraiman B~I.
\newblock Fluctuations of fitness distributions and the rate of muller's
  ratchet 2012 {\em Genetics\/} {\bf 191} 1283--1293

\bibitem{TKK2014}
Takeuchi N, Kaneko K and Koonin E~V.
\newblock Horizontal gene transfer can rescue prokaryotes from muller's
  ratchet: benefit of dna from dead cells and population subdivision 2014 {\em
  G3: Genes, Genomes, Genetics\/} {\bf 4} 325--339

\bibitem{OtwKru2014}
Otwinowski J and Krug J.
\newblock Clonal interference and muller's ratchet in spatial habitats 2014
  {\em Physical biology\/} {\bf 11} 056003

\bibitem{LAKDG2011}
Luke{\v{s}} J, Archibald J~M, Keeling P~J, Doolittle W~F and Gray M~W.
\newblock How a neutral evolutionary ratchet can build cellular complexity 2011
  {\em IUBMB life\/} {\bf 63} 528--537

\bibitem{DDST2015}
De~Vos M~G, Dawid A, Sunderlikova V and Tans S~J.
\newblock Breaking evolutionary constraint with a tradeoff ratchet 2015 {\em
  Proceedings of the National Academy of Sciences\/} {\bf 112} 14906--14911

\bibitem{WeiCha2005}
Weinreich D~M and Chao L.
\newblock Rapid evolutionary escape by large populations from local fitness
  peaks is likely in nature 2005 {\em Evolution\/} {\bf 59} 1175--1182

\bibitem{CroKim1970}
Crow J~F, Kimura M {\em et~al.\/}.
\newblock An introduction to population genetics theory. 1970 {\em An
  introduction to population genetics theory.\/}

\bibitem{Kim1962}
Kimura M.
\newblock On the probability of fixation of mutant genes in a population 1962
  {\em Genetics\/} {\bf 47} 713

\bibitem{Sei2012}
Seifert U.
\newblock Stochastic thermodynamics, fluctuation theorems and molecular
  machines 2012 {\em Reports on progress in physics\/} {\bf 75} 126001

\bibitem{DecSas2018}
Dechant A and Sasa S~i.
\newblock Current fluctuations and transport efficiency for general langevin
  systems 2018 {\em Journal of Statistical Mechanics: Theory and Experiment\/}
  {\bf 2018} 063209

\end{thebibliography}

\end{document}